# Imaginary Poynting momentum driven particle bidirectional rotation along arbitrary trajectory

LIFANG ZHAO [1], XUE YUN, YANSHENG LIANG[1,2], MING LEI[1,3]

[1] *MOE Key Laboratory for Nonequilibrium Synthesis and Modulation of Condensed Matter, Shaanxi Province Key Laboratory of Quantum Information and Quantum Optoelectronic Devices, School of Physics, Xi'an Jiaotong University, Xi'an 710049, China*

[2] *yansheng.liang@mail.xjtu.edu.cn.*

[3] *ming.lei@mail.xjtu.edu.cn.*

**Abstract:** Research on optical rotational manipulation leveraging the imaginary Poynting momentum (IPM) force has predominantly centered on cylindrically polarized Gaussian and annular beams. Here, we extend this framework to tightly focused cylindrically polarized structured light fields possessing closed or open arbitrary-intensity trajectories. We systematically elucidate the rotational dynamics induced by IPM in such fields, characterizing the underlying mechanical effects and trapping flexibilities. Notably, despite carrying zero net angular momentum, these fields drive microparticles into bidirectional rotation along predefined trajectories, thereby challenging conventional paradigms reliant on spin or orbital angular momentum. This IPM-mediated optical spanner offers unprecedented spatial degrees of freedom, paving the way for high-precision optical manipulation along arbitrarily tailored paths.

## 1. Introduction

Structured and complex-shaped beams have proven instrumental in elucidating optical forces and light–matter interactions underpinning optical micromanipulation [1-5]. Integrating such tailored fields into optical tweezers facilitates multifunctional particle control, substantially broadening the application scope of the field [6-9]. Traditionally, optical forces exerted on mesoscopic particles arise from intensity and phase gradients of light [10-13], wherein radiation pressure is intrinsically linked to the real part of the Poynting momentum [14]. In contrast, mechanical effects stemming from the imaginary part of the Poynting vector—the imaginary Poynting momentum (IPM)—have garnered significant attention only recently [15-18]. A pivotal conceptual breakthrough occurred in 2010 when Nieto-Vesperinas *et al.* first theorized that IPM, long dismissed as a mathematical artifact devoid of physical significance, could in fact induce measurable mechanical effects [19]. Initial observations of IPM-driven dynamics emerged in evanescent and interference fields [15, 16, 20, 21]. However, the transverse forces detected in these studies were not exclusively attributable to IPM. A major advance was reported by Xu *et al.*, who demonstrated that paraxial cylindrically polarized beams possessing purely azimuthal IPM density could drive orbital particle motion even in the absence of spin angular momentum (SAM) and orbital angular momentum (OAM) [22]. Subsequent refinements in theory have catalyzed successive experimental breakthroughs [23, 24]. Notably, Zhou *et al.* constructed an IPM vortex beam to achieve bidirectional rotation of gold nanoparticles on opposing sides of a circular trajectory, delivering the first unambiguous experimental evidence of the IPM effect [23]. Similarly, studies employing cylindrically polarized Gaussian beams revealed IPM-induced rotation of gold particles with polarization-tunable directionality [24]. Despite extensive exploitation of IPM in evanescent waves and structured light, experimental strategies for realizing IPM-driven rotation remain scarce. Arbitrary trajectory beams transcend the limitations of conventional ring-shaped intensity profiles, permitting particle transport along customized closed or open paths [25-29]. Exploiting IPM forces within these fields unlocks new possibilities for microparticle manipulation, specifically enabling rotational driving along arbitrarily complex, user-defined trajectories.

In this work, we investigate IPM-mediated particle rotation induced by tightly focused cylindrically polarized arbitrary-trajectory beams. Through combined theoretical simulations and experiments, we demonstrate that azimuthal forces arising from IPM transfer drive gold particles into bidirectional rotation on either side of the trajectory, even in the complete absence of SAM and OAM. Our findings advance the fundamental understanding of IPM dynamics and establish a robust pathway toward precision rotational manipulation along customizable trajectories.

## 2. Principle

Optical torque conventionally arises from the transfer of angular momentum, specifically spin angular momentum (SAM) or orbital angular momentum (OAM) [30-32]. Crucially, these torque-generating components are intrinsically linked to the real part of the complex Poynting vector, representing the time-averaged linear momentum density. The complex Poynting vector is defined as:

$$\mathbf{\Pi} = \frac{1}{2}\left(\mathbf{E}^* \times \mathbf{H}\right) \tag{1}$$

In contrast, its imaginary component—the imaginary Poynting momentum (IPM)—can induce substantial mechanical effects. Particularly when cylindrically polarized beams are tightly focused by a high numerical aperture (NA) objective, this IPM drives particles into orbital motion. Here, we elucidate the mechanism governing IPM-induced rotation in generic cylindrically polarized vector fields. The incident electric field in the Cartesian basis is expressed as:

$$\mathbf{E}_0(\rho,\varphi) = E_0(\theta)(\cos\alpha \mathbf{e}_\rho + \sin\alpha \mathbf{e}_\varphi) \tag{2}$$

Where $E_0(\theta)$ denotes the amplitude of the incident optical field with $\theta=\arcsin(\rho/f)$, and $\alpha$ is the polarization angle that determines the initial polarization state of the cylindrically polarized vector beam. The unit vectors in the polar coordinate system are denoted by $\mathbf{e}_\rho$ and $\mathbf{e}_\varphi$. Using the Richards-Wolf vectorial diffraction integral theory, the tight focusing electric field components near the focal plane are derived as[33-35]:

$$\begin{aligned}
E_r(r,\phi,z) &= -kf\cos\alpha \int_0^{\theta_{\max}} E_0\sqrt{\cos\theta}\sin\theta\cos\theta J_1(kr\sin\theta)\exp(ikz\cos\theta)d\theta \\
E_\phi(r,\phi,z) &= -kf\sin\alpha \int_0^{\theta_{\max}} E_0\sqrt{\cos\theta}\sin\theta J_1(kr\sin\theta)\exp(ikz\cos\theta)d\theta \\
E_z(r,\phi,z) &= -ikf\cos\alpha \int_0^{\theta_{\max}} E_0\sqrt{\cos\theta}\sin^2\theta J_0(kr\sin\theta)\exp(ikz\cos\theta)d\theta
\end{aligned} \tag{3}$$

where $k$ is the wavenumber and $\theta_{\max}$ denotes the maximal angle determined by the NA of the objective lens. To simplify the equation, we substitute the expression[23]

$$U(r,z) = -kf\int_0^{\theta_{\max}} E_0\sqrt{\cos\theta}\sin\theta J_1(kr\sin\theta)\exp(ikz\cos\theta)d\theta \tag{4}$$

into Eq. 3. Then we obtain the total electric field:

$$\mathbf{E}(r,\phi,z) = -i\frac{\cos\alpha}{k}\partial_z U(r,z)\mathbf{e}_r + U(r,z)\mathbf{e}_\phi + i\frac{\cos\alpha}{k}(\partial_r + \frac{1}{r})U(r,z)\mathbf{e}_z. \tag{5}$$

According to Maxwell's equations, the total magnetic field is obtained:

$$\mathbf{H}(r,\phi,z) = i\frac{\sin\alpha}{\mu ck}\partial_z U(r,z)\mathbf{e}_r + \frac{1}{\mu c}\cos\alpha U(r,z)\mathbf{e}_\phi - i\frac{\cos\alpha}{\mu ck}(\partial_r + \frac{1}{r})U(r,z)\mathbf{e}_z \tag{6}$$

where $\mu$ is the medium permeability and $c$ is the light speed in the medium. Therefore, according to Eq. (1), Eq. (5), and Eq. (6), the real part of the complex Poynting vector is

$$\begin{aligned}\mathrm{Re}\left(\mathbf{\Pi}\right) &= \mathrm{Re}\left(\Pi_r\right) + \mathrm{Re}\left(\Pi_\phi\right) + \mathrm{Re}\left(\Pi_z\right) \\ &= \frac{1}{2\mu ck}\mathrm{Im}\left(U^*\cdot\partial_r U\right)\mathbf{e}_r + \frac{1}{2\mu ck}\mathrm{Im}\left(U^*\cdot\partial_z U\right)\mathbf{e}_z\end{aligned} \tag{7}$$

This part represents time-averaged momentum density, with its angular component equal to zero. The IPM density is

$$\begin{aligned}\mathrm{Im}\left(\mathbf{\Pi}\right) &= \mathrm{Im}\left(\Pi_r\right) + \mathrm{Im}\left(\Pi_\phi\right) + \mathrm{Im}\left(\Pi_z\right) \\ &= \frac{\cos 2\alpha}{2\mu ck}\left(\frac{1}{r}|U|^2 + \frac{1}{2}\partial_r|U|^2\right)\mathbf{e}_r + \frac{\sin 2\alpha}{2\mu ck^2}\mathrm{Im}\left[\frac{1}{r}\partial_r(rU^*)\partial_z U\right]\mathbf{e}_\phi \\ &+ \frac{\cos 2\alpha}{2\mu ck}\mathrm{Re}\left(U^*\cdot\partial_z U\right)\mathbf{e}_z\end{aligned} \tag{8}$$

The rotational dynamics of the IPM for a generic cylindrically polarized vector beam is described by Equation (8), The IPM density depends on the polarization angle $\alpha$[24]. The polarization angle refers to the angle between the electric field direction and the radial direction in cylindrical vector beam, with a range of −90° to 90°. Among these, 0° corresponds to radial polarization, and ±90° correspond to azimuthal polarization.

As an example, Fig. 1 presents the complex Poynting vector distributions in the focal plane of the tightly focused cylindrically polarized square beams for $\alpha$=−45°, 45°, 22.5°, and 0°. The simulation sets wavelength $\lambda$=1064 nm and NA= 1.33. As shown in Fig. 1(a), the total intensity distribution for different polarization angles reveals a double trajectory pattern in the focal plane with a local minimum between the two intensity maxima, and the intensity at this local minimum increases as the polarization angle decreases. Fig. 1(b) shows the real part of the complex Poynting vector, Re(**Π**). Fig. 1(c) presents Re(**Π**) along the line $y$=0. For the focused optical fields at different polarization angles, Re(**Π**) is mainly distributed in the longitudinal direction Re($\Pi_z$), which excludes contributions from the SAM and OAM to the orbital rotation of particles. Fig. 1(d) depicts the IPM flow, with the inset showing a magnified view of the local IPM streamline distribution within the region (x: −3.7 to −3.58 μm, y: −0.06 to 0.06 μm). Fig. 1(e) illustrates the radial component Im($\Pi_r$) and azimuthal component Im($\Pi_\phi$) of the IPM. When $\alpha$=45°, Im($\Pi_\phi$) reaches its maximum, while Im($\Pi_r$) is zero. If the polarization direction is reversed, the direction of the IPM flow also reverses. When $\alpha$ decreases to 0°, Im($\Pi_\phi$) disappears, and Im($\Pi_r$) reaches its maximum; in this case, the optical field produces no azimuthal force.

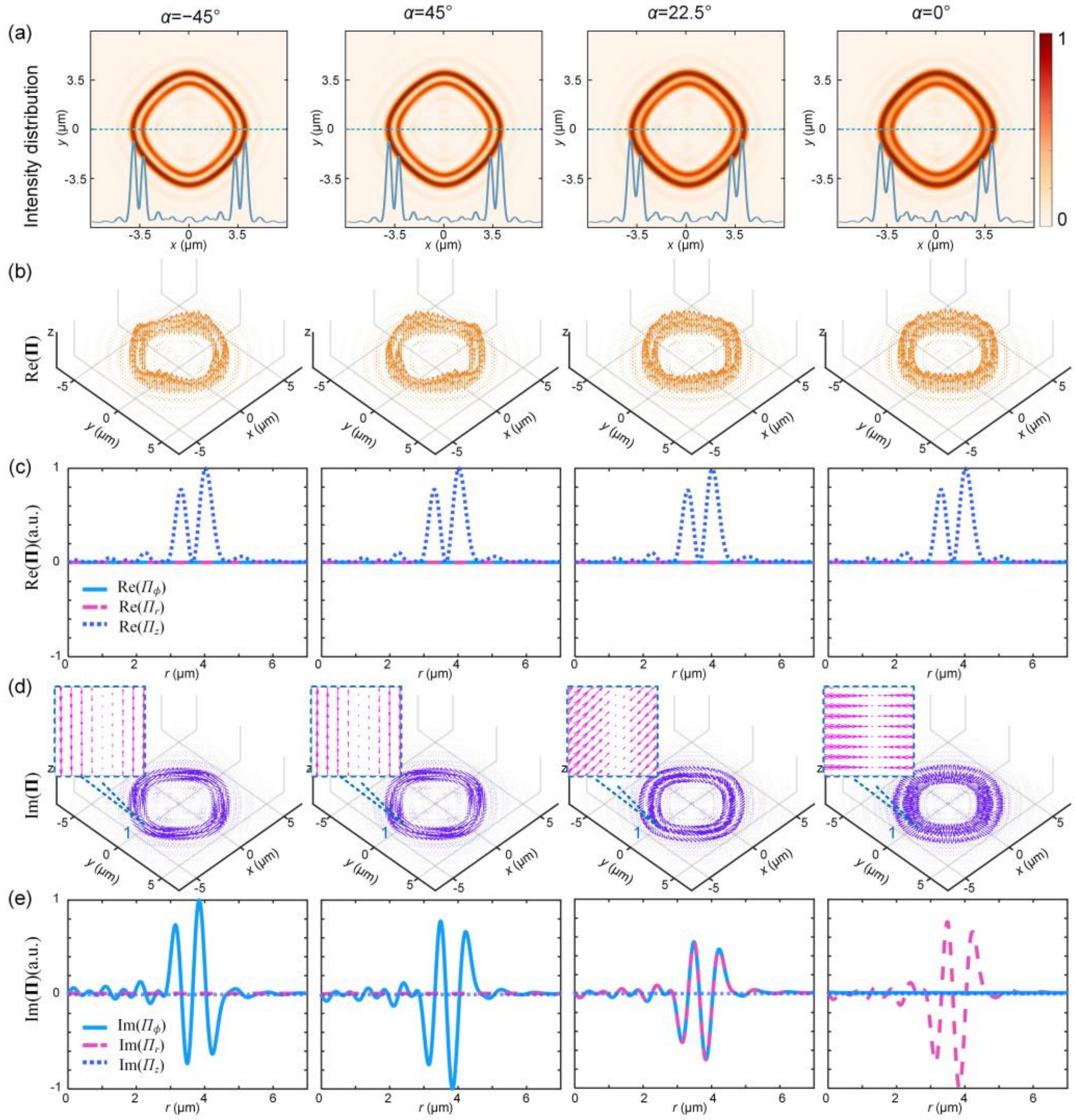

Fig. 1 IPM distribution of tightly focused cylindrical polarized square beams with different polarization angles. (a) The intensity distribution. (b)Re(**Π**). (c) The radial, azimuthal, and axial components of the Re(**Π**) related to the radial position. (d) Im(**Π**). (e) The radial, azimuthal, and axial components of Im(**Π**) related to the radial position.

## 3. Result

### A. Simulation Results

To analyze in detail the IPM distribution of cylindrically polarized arbitrary trajectory beams, Fig. 2 presents Re(**Π**) and Im(**Π**) of such fields with ring, ellipse, triangle, square, and pentagon shapes at polarization angle of −45°. Re(**Π**) exists only in the longitudinal direction, while the IPM lies in the xy-plane. Moreover, the IPM flow exhibits clearly opposite directions on the inner and outer sides of the trajectory field. Compared with ring optical fields, other arbitrary trajectory optical fields exhibit nonuniform momentum flow distribution along the trajectory at $\alpha$=−45°due to nonuniform curvature.

To observe the rotational effects induced by the IPM of tightly focused cylindrically polarized arbitrary trajectory beams, we numerically calculated the optical force characteristics of 4 μm diameter gold microparticles using the T-matrix method [36, 37]. In the numerical simulations, the NA of the entire system was set to 1.33, the optical power was set to 150 mW, and the complex refractive index of the gold microparticle at a wavelength of 1064 nm is 0.4 + 7.36i. Fig. 3 presents the optical force and trapping characteristics of gold microparticles in cylindrically polarized arbitrary trajectory beams with ring, triangle, square, and pentagon shapes at polarization angles of 45° and −45°. All net force distributions in the transverse plane (Fig. 3(b)) are concentrated on both sides of the geometric profiles of the optical field intensity (Fig. 3(a)). Fig. 3(c) presents the calculated radial and azimuthal forces along the radial direction. For beams with different geometric profiles, stable trapping positions are formed on both the inner and outer sides of the optical field (the inset in Fig. 3(c) shows a magnified view of the boxed region). At positions where the radial force $F_r$ is zero and its slope is negative, the directions of the azimuthal force $F_\phi$ on the two sides are always opposite. The force vector maps for regions 1 and 2 are shown in Fig. 3(d) and 3(e), respectively. On the two sides of the optical field trajectory, particles experience opposite driving forces, and when the sign of the polarization angle is reversed, the driving forces on the two sides also reverse. The calculated azimuthal forces at the equilibrium positions on both sides of the trajectory show that the inner forces are larger than the outer ones. For $\alpha$=−45°, the inner azimuthal forces for the ring, triangle, square, and pentagon are 0.9660 pN, 1.0328 pN, 1.0258 pN and 0.8511 pN, respectively, and the outer azimuthal forces are −0.7590 pN, −0.8325 pN, −0.6586 pN and −0.7632 pN, respectively. Reversing the polarization sign leaves the magnitudes unchanged but flips the directions.

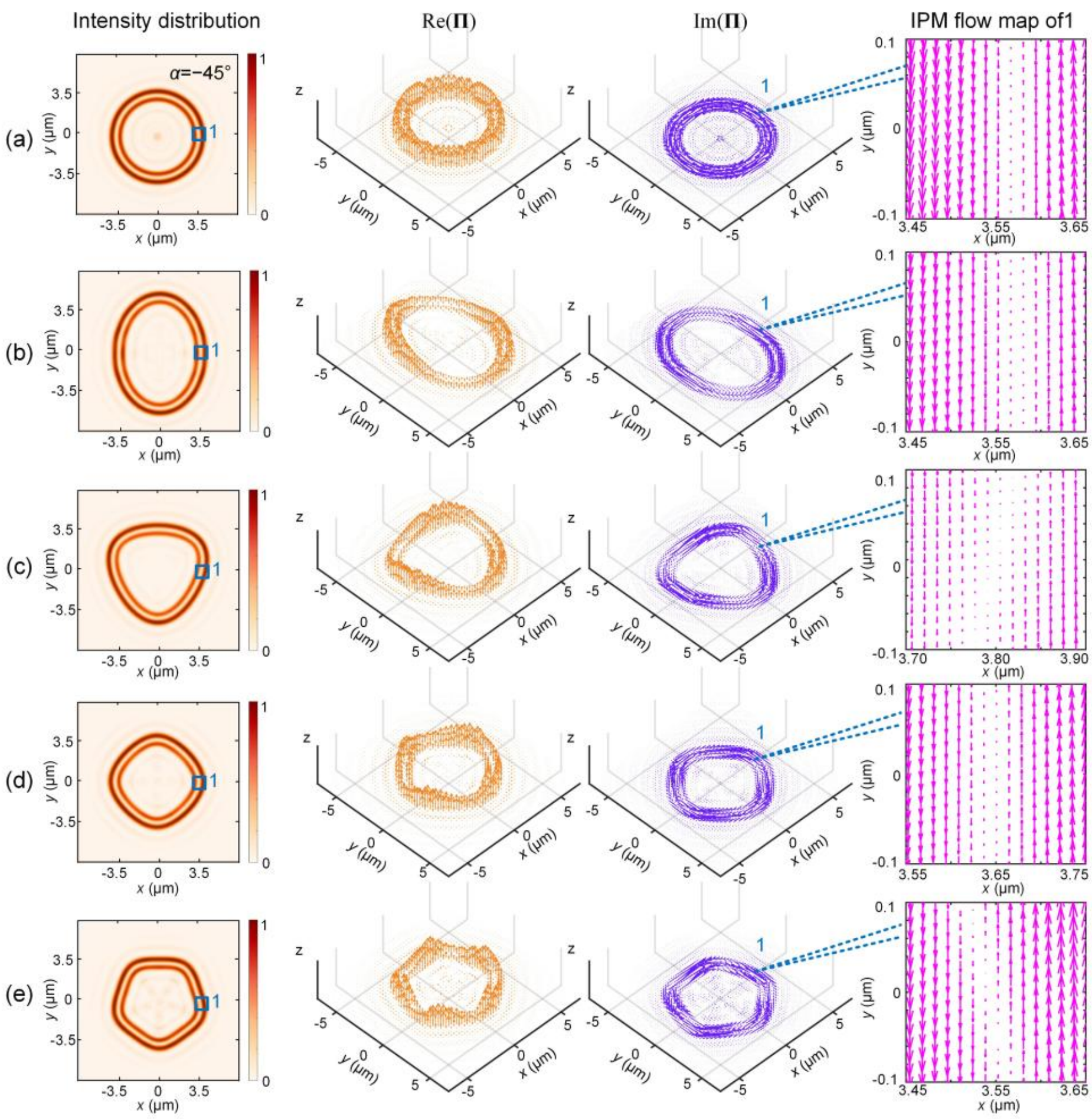

Fig. 2 Distributions of the real and imaginary parts of the complex Poynting vector for tightly focused cylindrical polarized arbitrary trajectory beams. (a)Ring (b)Ellipse (c)Triangle (d)Square (e)Pentagon

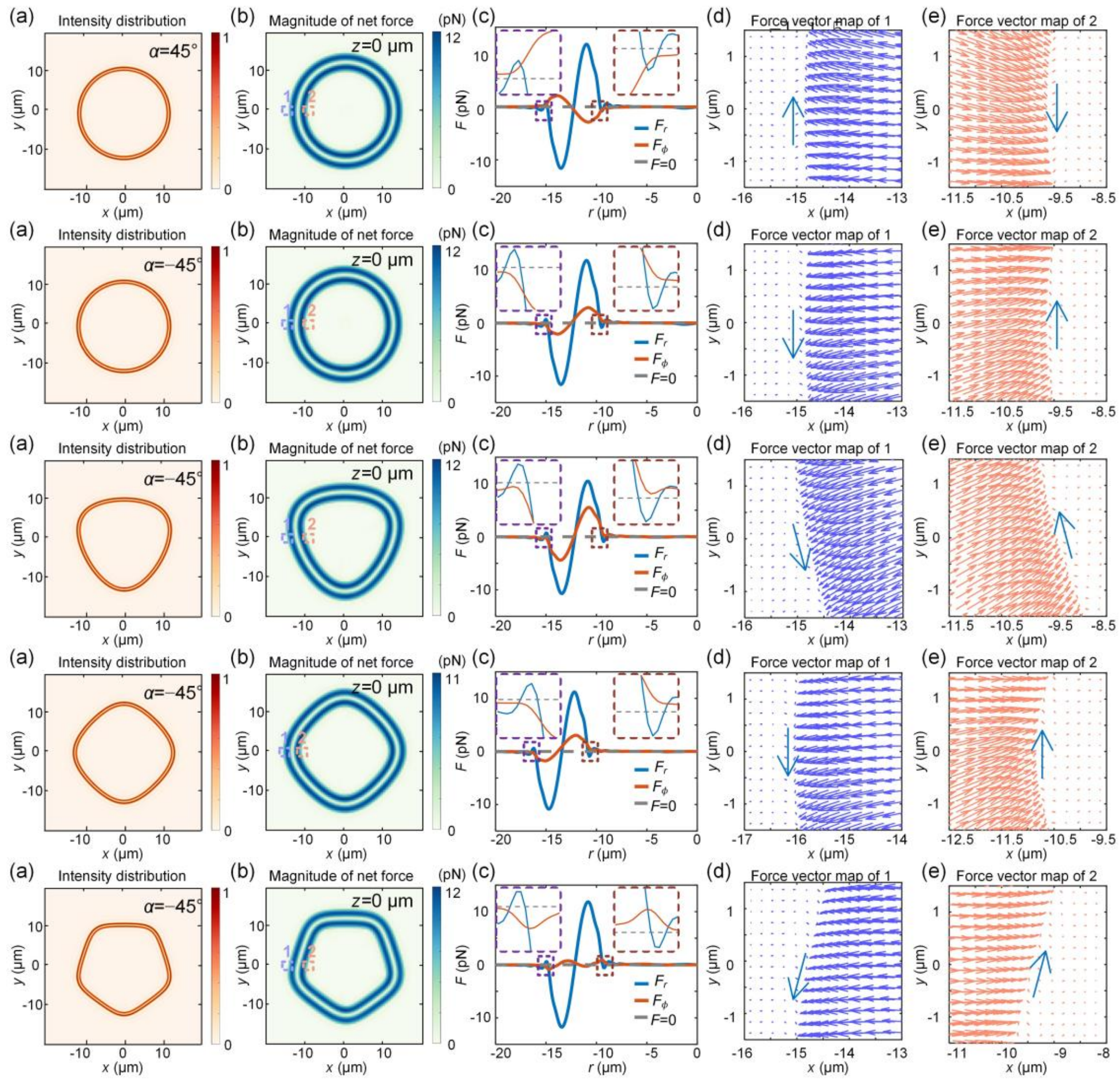

Fig. 3 Optical force and trapping characteristics for tightly focused cylindrically polarized arbitrary trajectory beams. (a) Intensity distribution. (b) Net force magnitude distribution. (c) Radial force $F_r$ and azimuthal force $F_\phi$ versus the radial position. (d) and (e) net force vector maps in regions 1 and 2.

Figure 4 presents the complex Poynting vector distributions of the cylindrical vector non-closed arbitrary trajectory beams and the optical force characteristics experienced by gold microparticles. Similar to the closed arbitrary trajectory beams, the real part of the Poynting vector is primarily distributed along the beam propagation direction, as shown in Fig. 4(b), whereas IPM is mainly confined to the transverse plane. For $\alpha$=−45°, the radial component of the IPM reaches its maximum value, and the IPM exhibits opposite directions on the two sides of the intensity minimum, as illustrated in Fig. 4(d). The net optical force distributions in the transverse plane for $\alpha$=−45° and $\alpha$=45° are shown in Fig. 4(e), 4(i), and 4(m), respectively. It can be observed that the optical force is predominantly localized on both sides of the optical trajectory, corresponding to regions with intensity gradients. In contrast, the optical force near the trajectory endpoints is significantly weaker, indicating that these regions cannot provide sufficient optical driving force. Consequently, stable trapping and directional transport mainly occur along the sidewalls of the trajectory rather than near its endpoints. The radial force $F_r$ and azimuthal force $F_\varphi$ calculated along the radial direction are presented in Fig. 4(f), 4(j), and 4(n). Stable equilibrium positions are formed on both the inner and outer sides of the trajectory, corresponding to locations where $F_r$= 0 and the force gradient is negative. Similar to the case of closed trajectories, particles trapped at these equilibrium positions experience azimuthal driving forces with opposite directions, resulting in directional transport along the prescribed trajectory. Fig. 4(g) and 4(h) show the force vector maps around equilibrium positions 2 and 3 for α = −45°, while Fig. 4(k), 4(l), 4(o), and 4(p) illustrate the force distributions around equilibrium positions 4 and 5 for α = −45° and α = 45°, respectively. It is evident that reversing the polarization angle from α = −45° to α = 45° completely reverses the direction of the azimuthal force, while the radial trapping characteristics remain unchanged. Quantitative analysis further reveals that the magnitudes of the azimuthal forces at the equilibrium positions on the two sides of the non-closed trajectory are unequal. For α = −45°, the azimuthal forces at equilibrium positions 4 and 5 are 1.5551 pN and −1.8310 pN, respectively. Reversing the sign of the polarization angle changes only the direction of the azimuthal force without affecting its magnitude. Therefore, the transport direction of the trapped particle can be controlled through polarization modulation.

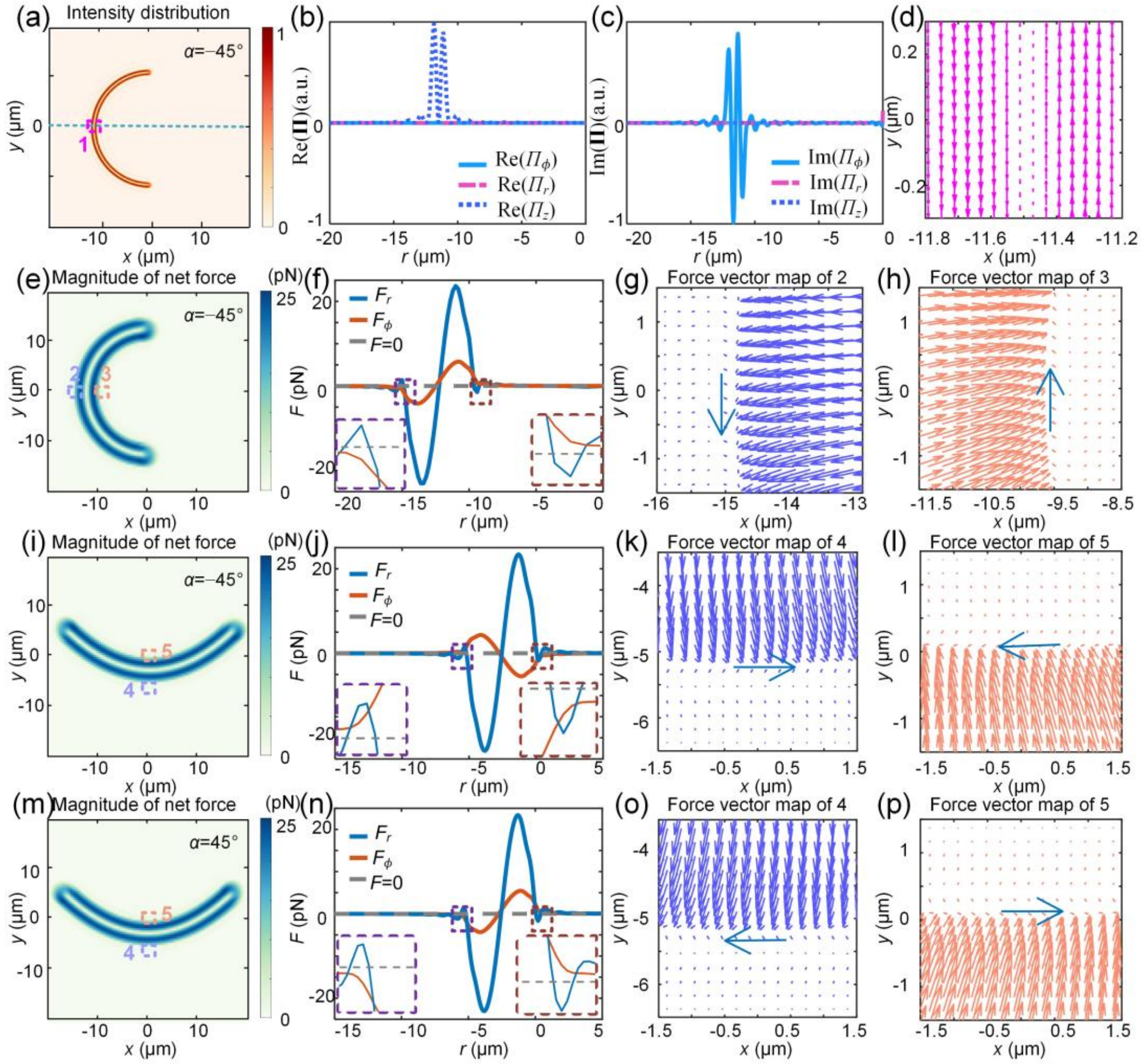

Fig. 4 Complex Poynting momentum distributions and optical force characteristics of gold nanoparticles in cylindrical vector non-closed arbitrary trajectory beams. (a) Intensity distribution. (b) Real part of the Poynting momentum. (c), (d) Imaginary part of the Poynting momentum. (e), (i), and (m) Net optical force distributions. (f), (j), and (n) Radial force $F_r$ and azimuthal force $F_\phi$. (g), (h), (k), (l), (o), and (p) Optical force vector maps at stable equilibrium positions.

## B. Experimental Results

Arbitrary-trajectory beams were synthesized via free-space lens modulation [29], and the corresponding experimental configuration is schematized in Fig. 5. A linearly polarized laser (λ=1064nm) was first expanded and collimated, then directed through a polarizing beam splitter (PBS) and a half-wave plate (HWP) to ensure a pure horizontal polarization state. A phase-only spatial light modulator (SLM) imprinted a hologram comprising a digital lens with a focal length of 200mm. A telescopic relay system (including lens L1) subsequently resized the beam to overfill the back aperture of the objective lens. Before entering the high-NA objective, the beam passed through a dichroic mirror and a vortex half-wave retarder (WPV10L-1064, Thorlabs), which converted the linearly polarized beam into the desired cylindrically polarized vector beam. The beam was then tightly focused onto the sample stage by a high-NA oil-immersion objective (100×, NA 1.45, Nikon). Imaging of the sample (illuminated by an LED) was performed via the same optical path: the back-scattered light transmitted through the dichroic mirror and was recorded by a CMOS camera (DCC3240M, Thorlabs). Particle trapping was achieved by positioning the sample slide at the focal plane using a three-axis piezoelectric stage.

Compared to conventional dual-lens tweezers systems, our design employs an SLM-generated digital lens to replace the subsequent physical Fourier lens. This integration significantly shortens the optical path, mitigates zero-order diffraction artifacts inherent to pixelated SLMs, and enhances the flexibility of beam steering. While theoretical models predict a distinct double-trajectory intensity profile at the focal plane, experimental constraints led to a deviation in the observed pattern. Specifically, the relatively large beam footprint required a second passage through the VHW prior to imaging, which distorted the intensity distribution away from the ideal theoretical prediction, as illustrated in Fig. 5.

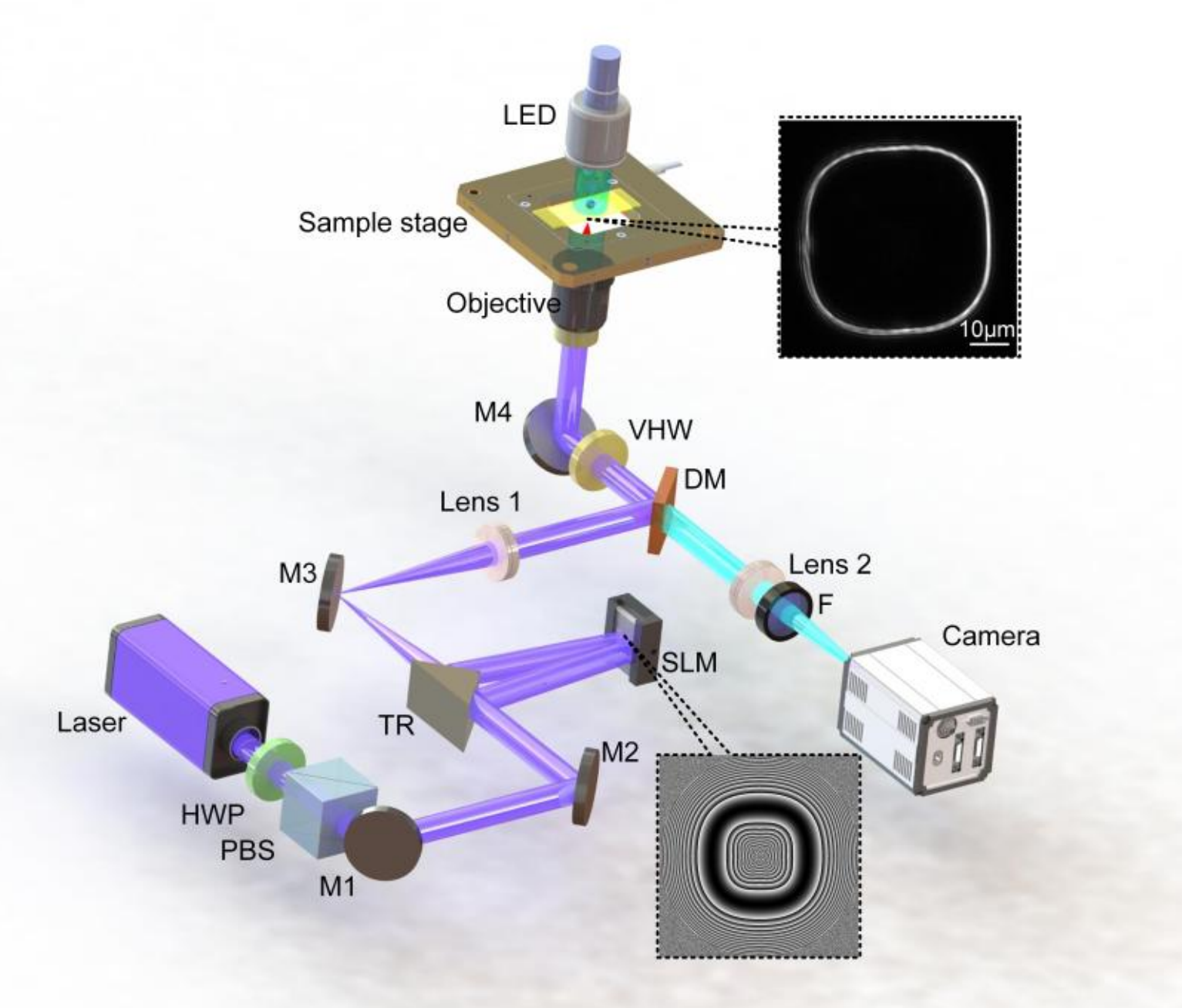

Fig. 5 Schematics of experimental set-ups. HWP, half-wave plate; PBS, polarizing beam splitter; M, mirror; SLM, spatial light modular; L, Lens; DM, dichroic mirror; VHW, vortex half-wave retarder. The polarization parameter $\alpha$ is controlled by VHW. The two insets show the hologram on the SLM and the intensity distribution at the focal plane, respectively.

Figure 6 presents the orbital rotation of two Au particles in tightly focused cylindrically polarized closed arbitrary trajectory beams. The experiment recorded video sequences of spherical Au particles with diameters ranging from approximately 3 to 5.5 μm moving along arbitrary trajectory beams of ring, triangle, square, and pentagon shapes. The two particles are radially confined to the outer and inner sides of the focal beam trajectory, respectively, and the IPM force acting in the azimuthal direction drives the particles to rotate. However, their rotational behaviors are distinctly different. At $\alpha$=45°, particle A on the outer side of the trajectory rotates clockwise, while particle B on the inner side rotates counterclockwise, resulting in opposite rotation directions. At $\alpha$=−45°, the rotation directions of both particle A on the outer side and particle B on the inner side reverse. These observations are in excellent agreement with the theoretical predictions, confirming that the azimuthal forces at the two radial equilibrium positions on opposite sides of the trajectory have opposite directions.

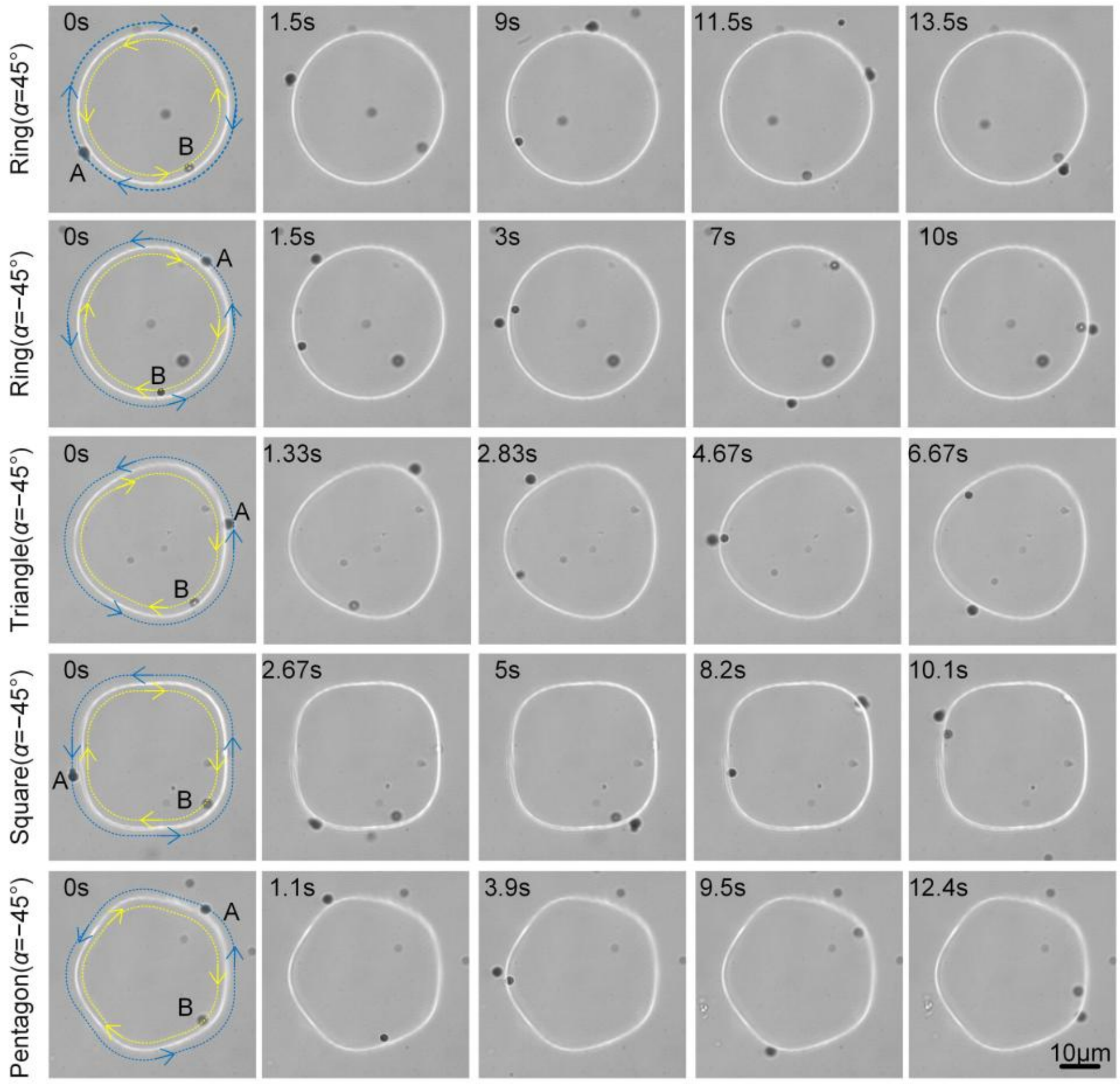

Fig. 6 Experimental results showing two Au particles trapped and rotated by tightly focused cylindrically polarized closed arbitrary trajectory beams. The blue dashed arrows and the yellow dashed arrows indicate the rotation trajectories and directions of the outer particle and the inner particle, respectively.

Figure 7 presents the experimental results of an Au particle motion in tightly focused cylindrically polarized non-closed arbitrary trajectory beams at $\alpha$=−45°. The Au particle is trapped at different radial equilibrium positions on the two sides of the optical trajectory and driven by azimuthal force. Particle trapped on opposite sides of the beams move along opposite directions, as indicated by the yellow and blue arrows. These experimental observations are consistent with the theoretical results presented in Fig. 4, demonstrating the IPM can simultaneously provide stable trapping and directional transport in non-closed arbitrary trajectory beams. In summary, for cylindrically polarized arbitrary trajectory beams (including both closed and non-closed trajectories), Au particles can be simultaneously and stably trapped on both the inner and outer sides of the trajectory and driven to move directionally along the trajectory, with the azimuthal forces acting on the particles on the two sides always opposite in direction.

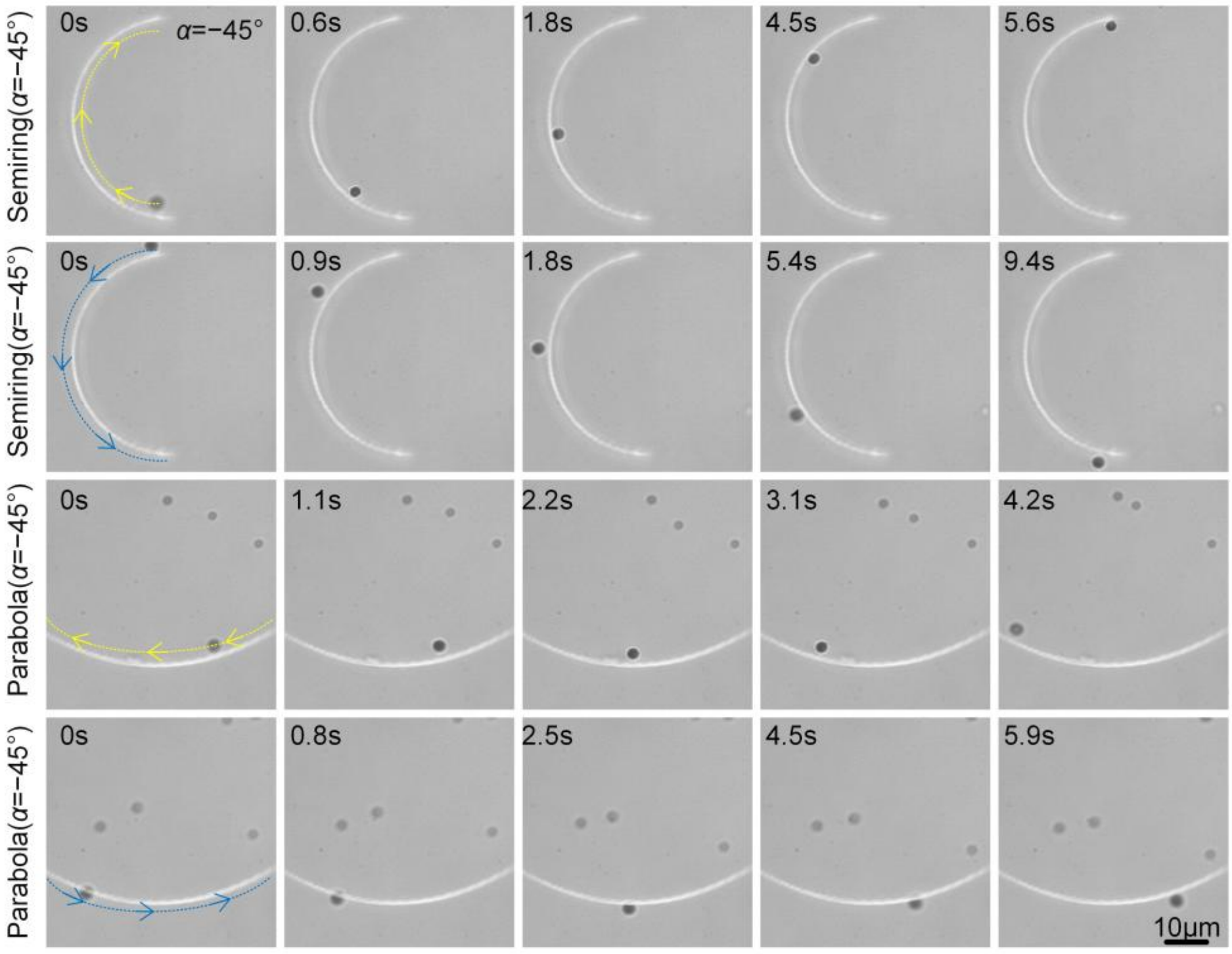

Fig. 7 Experimental results showing an Au particle trapped and moved by tightly focused cylindrically polarized non-closed arbitrary trajectory beams at $\alpha$=−45°. The directions of particle motion are indicated by blue and yellow arrows, respectively.

## 4. Conclusion

In summary, we have theoretically and experimentally demonstrated that tightly focused arbitrary-trajectory beams can induce robust orbital rotation of gold microparticles via the imaginary Poynting momentum (IPM) force. Remarkably, this bidirectional rotational control is achieved solely through IPM transfer, operating independently of conventional spin or orbital angular momentum. This capability establishes IPM as a novel degree of freedom in optical micromanipulation. Furthermore, our findings advance the fundamental understanding of momentum transfer mechanisms in complex structured light fields and lay the experimental groundwork for realizing reconfigurable bidirectional optical spanners along user-defined trajectories.